\pdfoutput=1
\documentclass[aip,jcp,reprint,floatfix]{revtex4-2}

\usepackage{amsmath}
\usepackage{graphicx}
\usepackage{hyperref}

\begin{document}


\title{Contrasting Mechanisms for Photodissociation of Methyl Halides Adsorbed on Thin Films of \texorpdfstring{C\(_6\)H\(_6\)}{C6H6} and \texorpdfstring{C\(_6\)F\(_6\)}{C6F6}} 



\author{E.T. Jensen}
\email[email:]{ejensen@unbc.ca}
\affiliation{Department of Physics \\University of Northern British Columbia, 3333 University Way, Prince George B.C. Canada V2N 4Z9}


\date{January 15, 2021}

\begin{abstract}
The mechanisms for photodissociation of methyl halides (CH$_3$X, X= Cl, Br, I) have been studied for these molecules when adsorbed on thin films of C$_6$H$_6$ or C$_6$F$_6$ on copper single crystals, using time-of-flight spectroscopy with 248nm and 193nm light. For CH$_3$Cl and CH$_3$Br monolayers adsorbed on C$_6$H$_6$, two photodissociation pathways can be identified-- neutral photodissociation similar to the gas-phase, and a dissociative electron attachment (DEA) pathway due to photoelectrons from the metal. The same methyl halides adsorbed on a C$_6$F$_6$ thin film display only neutral photodissociation, with the DEA pathway entirely absent due to intermolecular quenching via a LUMO-derived electronic band in the C$_6$F$_6$ thin film. For CH$_3$I adsorbed on a C$_6$F$_6$ thin film, illumination with 248nm light results in CH$_3$ photofragments departing due to neutral photodissociation via the A-band absorption. When CH$_3$I monolayers on C$_6$H$_6$ thin films are illuminated at the same wavelength, additional new photodissociation pathways are observed that are due to absorption in the molecular film with energy transfer leading to dissociation of the CH$_3$I molecules adsorbed on top. The proposed mechanism for this photodissociation is via a charge-transfer complex for the C$_6$H$_6$ layer and adsorbed CH$_3$I.
\end{abstract}

\maketitle 


\section{Introduction}
\label{introduction}

The need for understanding of the mechanisms for photochemical processes in heterogeneous molecular environments arises from a variety of areas of application, including astrochemistry and planetary sciences\cite{Arumainayagam:2019dl}, radiation chemistry\cite{Alizadeh:2012bz} and molecular devices\cite{Adams:2003kh,Zhu:2004jr}. The application of the tools of surface science to study these processes in well-characterized systems allows details of such processes to be laid out. There have been relatively few studies of photodissociation dynamics in these types of heterogeneous systems. In previous studies we have characterized near-UV photochemical processes for methyl halides adsorbed on D$_2$O and CH$_3$OH ices on metal substrates. The present work extends this to two small aromatic molecules, benzene and its fluorinated counterpart perfluorobenzene. We have studied a range of methyl halides (CH$_3$X, X= Cl, Br, I) adsorbed on thin films of C$_6$H$_6$ or C$_6$F$_6$ on Cu single crystal substrates. The stimulated dissociation properties of these methyl halides have been studied in some detail in both the gas-phase\cite{Eppink:1998ue,Gougousi:1998to,Townsend:2004uu,Wang:2014vj} as well as condensed on surfaces\cite{Marsh:1988vm,Ayotte:1997th,Jensen:2005js,Jensen:2008jv,Jensen:2015dg}. Depending on the context, these molecules can display low-energy photoelectron driven Dissociative Electron Attachment (DEA) or neutral photodissociation processes when in the adsorbed state, with outcomes dependent on the details of the particular molecule, intermolecular interactions and electronic structure of the environment.

\subsection{Adsorption and Valence Band Structure of \texorpdfstring{C$_6$X$_6$}{C6X6} on Cu Surfaces}
On both the Cu(110) and Cu(100) substrates C$_6$H$_6$ is believed to grow in a flat first monolayer with an upright second layer\cite{Dougherty:2006jl} due to the quadrupole moment of C$_6$H$_6$ favouring a T-motif in the intermolecular interaction. For thicker C$_6$H$_6$ films, it is believed that a herringbone structure is dominant\cite{Lee:2006gq}, similar to that seen in bulk crystals of the solid. Given that C$_6$F$_6$ has a quadrupole moment of similar magnitude though with opposite sign, it is at first surprising that C$_6$F$_6$ layers have been found to grow in a planar fashion\cite{Zhao:2014ia}\cite{Vijayalakshmi:2006ka}. This is due to the electrostatic interaction favouring an F-atom pointing to the centre of the carbon bond of a neighbouring molecule, and C$_6$F$_6$ films are believed to grown in this flat layer-by-layer structure for multilayer films. 

The  charge distribution and intermolecular interactions for adsorbed C$_6$F$_6$ leads to intermolecular \(\sigma\)-bonding to form a LUMO-derived unoccupied surface bandstructure\cite{Dougherty:2012jy}. The resultant $\sigma^*$ derived band is delocalized and in thin films displays nearly free electron-like dispersion. This LUMO-derived band of C$_6$F$_6$ on Cu surfaces forms an electronic quantum well state that shifts slightly in energy as the thickness of the C$_6$F$_6$ adlayers is varied\cite{Lindstrom:2006th,Zhao:2014ia}. This LUMO-derived band of the C$_6$F$_6$ thin film can also hybridize with the underlying substrate surface states at similar energies, leading to different interactions on Cu crystal substrates as a consequence of the presence or absence of surface bandgaps in the relevant range of energy. For this reason, we studied the photochemical behaviour for the methyl halides on C$_6$F$_6$ on Cu(110) (lacking a bandgap in the zone centre between E$_F$ and E$_{vac}$) and also that for the C$_6$F$_6$ on Cu(100) which has a substantial bandgap in this region\cite{Grass:1993dg}.  

In contrast, C$_6$H$_6$/Cu(111) has been found to have its LUMO \(\pi^*\) state at $E_F$+4.6eV observed for the bilayer\cite{Velic:1998vf,Frank:1986aa}, above the vacuum level, at a notably higher energy than the C$_6$F$_6$ LUMO. Films of C$_6$H$_6$ have been studied by optical spectroscopies-- for monolayer C$_6$H$_6$ on Cu\cite{Peng:2000up}, and for thicker C$_6$H$_6$ films grown on MgF$_2$\cite{Dawes:2017dpa} and on HOPG\cite{Stubbing:2020fy}. The near-UV absorption bands are observed to be shifted and broadened, with other modifications noted in the solid phase as compared to the gas-phase spectrum. A series of vibronic features due to ${^1A_{1g}}\rightarrow{^1B_{2u}}$ excitation lie in the energy region of relevance in the present work.

\subsection{Dissociation Mechanisms for Methyl Halides}
\label{CH3X_pdissn}
Photodissociation of gas-phase CH$_3$I in the near-UV region is dominated by the `A-band', a set of $n\rightarrow\sigma^*$ transitions (from the lone pair on the halogen to a C--I antibonding orbital) observed as three overlapping states ($^3Q_1$, $^3Q_0$ and $^1Q_1$ in order of increasing energy) in the Franck-Condon region\cite{Eppink:1998ue}. At the 248nm wavelength used in the present work, the $X\rightarrow {^3Q_0}$ excitation dominates and the $X\rightarrow {^1Q_1}$ is a minor channel. The subsequent dissociation can proceed via two principal pathways:
\begin{equation} \label{Equ_1}
\begin{split}
CH_3I + h\nu & \rightarrow CH_3 + I(^2P_{3/2}) \textrm{ \{ground state I\}} \\
& \rightarrow CH_3 + I^*(^2P_{1/2})  \textrm{ \{spin-orbit excited I\}}
\end{split}
\end{equation}

The energy difference between ground state $I$ and excited $I^*$ is 0.943eV, leading to significant differences in the translational energies imparted to the fragments and which can be resolved in our time-of-flight measurements. There are also vibrational and rotational energy partitioning differences for the CH$_3$ photofragments along the two pathways. Another significant factor for this system is that the $X-{^3Q_0}$ excitation is a parallel transition (requiring a component of the incident $\vec{E}$-field along the C--I bond axis), while the $X-{^1Q_1}$ excitation is perpendicular. This polarization dependence for optical absorption leads to being able to utilize polarization and molecular orientation to aid in understanding the photodissociation dynamics at 248nm\cite{Jensen:2005js}. The $^3Q_0$ state correlates to the I$^*$ outcome in Equ.~{\ref{Equ_1}}, but a curve-crossing with the $^1Q_1$ state (which correlates to the I pathway) during dissociation enables non-adiabatic transitions that result in both pathways being observed in experiments\cite{Eppink:1998ue}. 

Gas-phase photodissociation of CH$_3$Br\cite{Gougousi:1998to,Wang:2014vj} and CH$_3$Cl\cite{Townsend:2004uu} can also occur via the A-band, similar to the situation outlined for CH$_3$I above but at slightly higher energies. In contrast to the case for CH$_3$I at 248nm, the 193nm photodissociation of CH$_3$Br and CH$_3$Cl is dominated by a perpendicular transition from the ground state, mainly to the $^1Q_1$ state, which correlates to dissociation to CH$_3$ and a ground-state halogen atom. For CH$_3$Br  in the gas-phase, about 30\% (for CH$_3$Cl, 13\%) of the initial excitation is to the $^3Q_0$ state and there is also evidence\cite{Wang:2014vj,Townsend:2004uu} for the non-adiabatic coupling at the curve-crossing between the $^1Q_1$ and $^3Q_0$ states, leading to the CH$_3$ and spin-orbit excited Br$^*$ (or Cl$^*$) as outcomes. However, the smaller spin-orbit energies for Br and Cl do not allow resolvable TOF features for these pathways in the present work and so the changes in dynamics seen for different light polarizations are not as useful for CH$_3$Br and CH$_3$Cl as it is for CH$_3$I in the A-band. 

Dissociative electron attachment is well known for halomethanes in the gas-phase and has been observed for these molecules in the condensed phase in a variety of contexts. Of primary interest are the lowest energy DEA resonances, as these are often the pathway for low-energy photoelectrons or secondary electron cascades to cause dissociation in the condensed phase. A significant feature of condensed phase DEA is that the anionic dissociative state is shifted to lower energy relative to neutral states-- a consequence of dielectric screening of the anion due to the proximity of the metal surface and the dielectric response of the molecular environment in the thin film. This energetic shift not only reduces the resonance (attachment) energy but more significantly, reduces the time for the molecular bond lengthening to cross the "point of no return" beyond which the dissociative anionic state is energetically below the bound neutral one. For many halomethanes, the consequence of this is a substantially increased cross section for DEA in the adsorbed state-- in some cases by orders of magnitude (for example, for CH$_3$Cl by a factor of $10^4-10^6$)\cite{Ayotte:1997th}. Other factors can play a role in the net DEA cross section. One of pertinence for the present work is that for CH$_3$I, the low energy DEA cross sections are suppressed in the condensed state compared to the gas-phase\cite{Jensen:2008jv}, due to a disruption of the long-range electron-molecule interactions that lead to capture the incident electron, in this case a vibrational Feshbach resonance\cite{Fabrikant:2011ep}. The photon energies used in the present work are below the ionization thresholds for the molecules being studied. The principle source of low-energy electrons is the metal near-surface region, where photon absorption leads to hot photoelectrons (energies between $E_F$ and $E_{vac}$) and photoelectrons (energies above $E_{vac}$ if $h\nu > \Phi$, the workfunction)\cite{Weik:1993wa} can be generated and transported through intervening molecular thin film layers. If the photoelectrons initiate DEA, we refer to this process as Charge-Transfer or CT-DEA.

For either A-band neutral photodissociation or CT-DEA of the halomethanes, the dissociation process proceeds rapidly, with bond-breaking occurring in a few tens of femtoseconds. Photodissociation of halomethanes adsorbed on or close to a metal surface can be inhibited by quenching\cite{Zimmermann:1995wz,Zhou:1995}. When several layers of halomethanes are adsorbed, or are adsorbed on top of a spacer layer of another species, both CT-DEA and neutral photodissociation have been observed. Quenching of one or both photodissociation pathways by the surface at these timescales requires a rapid interaction, such as resonant electron/hole transfer between the excited molecule and the substrate\cite{Zhou:1995,Lindstrom:2006th}.

\subsection{Energetics of Stimulated Dissociation}
The dissociation of a CH$_3$X molecule in free space requires momentum and energy conservation, which determines how the excess kinetic energy is partitioned between the CH$_3$ fragment and the halogen atom. For a CT-DEA process the following can be used to rationalize the CH$_3$ photofragment kinetic energy in terms of the component factors:

\begin{eqnarray}
T_{{CH_3}}
& =  \frac{m(X)}{m(CH_3X)} \{& E_{e^-}+EA(X) - D_0(C-X)  \nonumber \\
& & + \Delta E_{solv}(X^-) - E_{int}(CH_3)  
\label{Equ_2}\}
\end{eqnarray}


where $m()$ is the mass of the particular species, $E_{e-}$ is the incident electron energy, $EA(X)$ is the electron affinity for the halogen atom $X$, $D_0$ is the energy of the bond being broken, $\Delta E_{solv}(X^-)$ is the energy of solvation for the product anion in its dielectric environment and $E_{int}(CH_3)$ is the internal energy (vibration and rotation) of the departing methyl fragment. In principle the solvation energy can be estimated\cite{Sun:1995vm,Marinica:2001ii} but the uncertainty in various parameters leads to $\Delta E_{solv}$ values that have fairly large uncertainty. This is particularly true in the heterogeneous molecular environments of dipolar molecules that we are considering in the present work, in which the solvation energy is structure and site sensitive, and can shift dynamically as the dissociation proceeds. The electron attachment energy $E_{e^-}$ is selected from the range of photoelectron energies created by the incident photons at the metal-molecule-vacuum interface\cite{Weik:1993wa}, (i.e. between the Fermi energy $E_F$ and $E_F + h\nu$) of which a portion will correspond to the attachment resonance energy of the molecule. For the methyl halides in the present work, it is believed that the peak of the DEA resonances of interest are near or below the vacuum level.

For neutral photodissociation, the analogous equation for the CH$_3$ photofragment kinetic energy is:

\begin{eqnarray}
T_{CH_3}
& =\frac{m(X)}{m(CH_3X)} \{ &h\nu - D_0(C-X) -  E_{int}(X) \nonumber \\
& & - E_{int}(CH_3) \}
\label{Equ_3}
\end{eqnarray}


where $h\nu$ is the photon energy, and $E_{int}(X)$ allows for the possible electronic excitation of the departing halogen atom. 

In surface systems the parent molecule is not in free space, but embedded at or near the vacuuum interface of the system being studied. It is known from prior work in gas-phase cluster and surface photochemistry that the observed fragment kinetic energy distributions can be altered by chemical or post-dissociation interactions, however Eqs.~{\ref{Equ_2}} and {\ref{Equ_3}} provide a basis to begin consideration of the observed kinetic energy distributions.

\section{Experimental Details}
The experiments were performed in an ultra-high vacuum (UHV) system that has been described previously\cite{Jensen:2005js}. The single crystal samples were cooled by liquid nitrogen (base temperature 90K) and heated by electron bombardment to 920K for cleaning. Sample temperatures were monitored by a type K thermocouple spot-welded to the tungsten sample mounting wire. Sample cleanliness and order were monitored by Auger electron spectroscopy (AES) and low energy electron diffraction (LEED) measurements respectively. Single crystals of Cu(110) and Cu(100) were used in this work. The crystals are 12mm diameter and were prepared in UHV by cycles of Ar$^+$ ion bombardment and electron bombardment heating and annealing until the sample AES spectra indicated clean copper substrates and the LEED patterns were of a ($1\times 1$) surface.

Temperature programmed desorption (TPD) measurements were made by rotating the sample to face a quadrupole mass spectrometer (QMS; UTI 100C) with its ionizer 76mm away, and heating the sample using the electron filament located a few mm behind the sample mount.

The photodissociation experiments were performed using a second QMS (Extrel). Neutral products from surface photodissociation travel 185mm to pass through a 4mm diameter aperture to a differentially pumped region with an axial electron bombardment ionizer. The sample to ionizer distance is 203mm. Ions created in the ionizer then travel through the quadrupole region and are mass selected, in the present experiments using m/q=15amu. Ion arrivals are recorded using a multichannel scaler that begins counting 50$\mu s$ prior to the initiating laser pulse, and the counts recorded from multiple laser pulses are summed. Unless otherwise indicated, the spectra shown in the present work are the result of summing data from 1000 laser pulses into 1000 1$\mu s$ time bins. In order for the ion arrival times to reflect the neutral fragment time-of-flight, they are corrected for the ion flight time (for CH$_3^+$, 17$\mu s$ at the 50eV ion energy used in the QMS). This is the leading systematic uncertainty in the recorded flight times ($\pm 1.5\mu s$) which does not affect comparisons between different TOF spectra but does lead to fixed nonlinear systematic uncertainty in the reported fragment kinetic energies $(KE\propto 1/(TOF)^2)$, which is most problematic at short flight times. The TOF spectra $N(t)$ were converted to probability distributions $P(E)$ versus CH$_3$ kinetic energy using the Jacobian transformation with a correction factor $1/t$ to account for the higher ionization probability for slower neutral CH$_3$ fragments\cite{Zimmermann:1995wz}. 

The laser pulses ($\sim$5ns duration) are produced by a small excimer laser (MPB PSX-100) operating at 20Hz. In this work  KrF ($\lambda$=248nm, $h\nu$=4.99eV) and ArF ($\lambda$=193nm, $h\nu$=6.42eV) laser light was used, with laser fluences on the sample of $\sim$ 0.8mJ/cm$^2$ or less per pulse, depending on the wavelength used. The intrinsic bandwidth of the laser emission for excimer lasers is rather broad. For a free-running KrF excimer laser the center wavelength is approximately 248.4nm (4.992eV) and has a fwhm bandwidth of $\sim$0.40nm (0.008eV). A possible implication of this property will be discussed in Section {\ref{sect_ch3I_c6h6}}. 

Both unpolarized and linearly polarized laser light has been used in this work for the reasons described in Section {\ref{CH3X_pdissn}}. When polarized light was used in the data presented, it is so indicated. In general the data obtained using 193nm light was unpolarized, though several measurements were repeated using polarized light with no differences aside from yield (peak heights) noted. To create polarized light, the beam passes through a birefringent MgF$_2$ crystal prism to separate p- and s-polarized components, which can then be directed at the sample.\footnote{For work at 248nm, s-polarized light was derived from the p-polarized beam by inserting an antireflection coated zero order half-waveplate into the beam. For 193nm, s-polarized light was obtained by rotating the MgF$_2$ prism to direct the s-polarized beam onto the sample.} The laser pulses were collimated using a 6mm diameter aperture and were unfocused on the sample. The laser light is incident upon the sample at a fixed angle of 45$^\circ$ from the TOF mass spectrometer axis-- for example, when the Cu crystal sample is oriented to collect desorption fragments along the surface normal direction, the light is incident at 45$^\circ$.

Cross sections for a selection of the molecular thin films examined in this work were determined by the depletion of the CH$_3$ photofragment yields. These ``depletion cross sections'' are obtained by recording CH$_3$ photofragment yields from photodissociation for a sequence of TOF spectra. Unpolarized laser light was used for these measurements. Time-of-flight spectra are obtained using 200--400 laser pulses per scan, then repeated for 10 or more successive scans. In the systems studied here, the TOF yields are observed to diminish as the net laser photon flux was increased, and the resulting yield vs. photon flux curves could be fit by a simple exponential decay model. Reasonable fits to the data were obtained, at least in the low flux limit. This procedure does not exclude the possibility that other photochemical processes involving the methyl halide but not seen in the TOF data might be occurring in the heterogeneous thin films. The reported cross sections have fairly large absolute errors (we estimate $\pm50\%$), but the errors in comparative relative cross sections between the different systems is much lower (10--20\% based on repeatability of measurements).

Deposition of molecules on the sample is done using a custom micro-capillary array directed doser based on the design of Ref.~\onlinecite{Fisher:2005uw}, with the sample held normal to the doser, 25mm away. This arrangement was found to enhance the deposition by a factor of 10 compared to background dosing. The pressure in the UHV chamber was measured using uncorrected ionization gauge readings. The dosing (in Langmuirs, L) was calibrated in terms of equivalent monolayers for the different species used by Temperature Programmed Desorption (TPD) measurements as discussed in Section {\ref{TPD}} below. The CH$_3$Br (Aldrich, $\ge$99.5\%) and CH$_3$Cl (Aldrich, $\ge$99.5\%) gases used in this work were transferred via a glass and teflon gas-handling system. The CH$_3$I (Aldrich, 99.5\%), C$_6$H$_6$ (Aldrich, 99.8\%) and C$_6$F$_6$ (Aldrich, 99.5\%) liquids used in this work were degassed by multiple freeze-pump-thaw cycles and the liquid contained in a pyrex vial a few cm from the precision leak valve used to admit the room-temperature vapour to the directed doser. 

\section{Results and Observations}
\subsection{Temperature Programmed Desorption}
\label{TPD}
In order to characterize the adsorption systems studied in this work, the thermal desorption of CH$_3$X (X=Cl, Br, I) and C$_6$X$_6$ (X=H, F) from Cu(110) and Cu(100) was studied. In the case of C$_6$H$_6$ adsorbed on Cu(110) and Cu(100), the features seen in TPD are consistent with previously published results\cite{Lee:2006gq}. The TPD measurements show that a dose of 0.35L of C$_6$H$_6$ corresponds to completion of the first monolayer. In the case of C$_6$F$_6$ adsorption, there are no published TPD results available for the Cu(110) or Cu(100) surfaces but our observations are broadly similar to those for C$_6$F$_6$/Cu(111)\cite{Vondrak:1999ks}. On Cu(111) the first layer desorption peak was seen at 193K while in our measurements we find that the peak of the first monolayer desorption is at 207K on Cu(110) and 203K on Cu(100). In both cases the monolayer equivalent dose is found to be 0.45L. We also find that the second layer desorption feature can be discerned at 167K with the multilayer desorption peak at 163K. For the methyl halides, the TPD results on the bare Cu surfaces have been characterized previously. We were able to perform TPD for CH$_3$Br on 2ML C$_6$F$_6$ and distinguish between the first layer desorption (129K) and multilayer desorption at 118K. The equivalent to monolayer CH$_3$Br dose on C$_6$F$_6$ is the same as that seen on the bare Cu substrates, within the experimental error of our measurements. A similar attempt for CH$_3$Cl on C$_6$F$_6$ could not distinguish the monolayer from multilayers (at 110K), so we have used the equivalent monolayer dose as determined on the bare Cu substrates.

\subsection{Photodissociation of \texorpdfstring{CH\(_3\)Cl}{CH3Cl} and \texorpdfstring{CH\(_3\)Br}{CH3Br} on \texorpdfstring{C\(_6\)H\(_6\)}{C6H6} and \texorpdfstring{C\(_6\)F\(_6\)}{C6F6} Thin Films}
\label{CH3X_results}
\begin{figure}[b]
\includegraphics[scale=0.55]{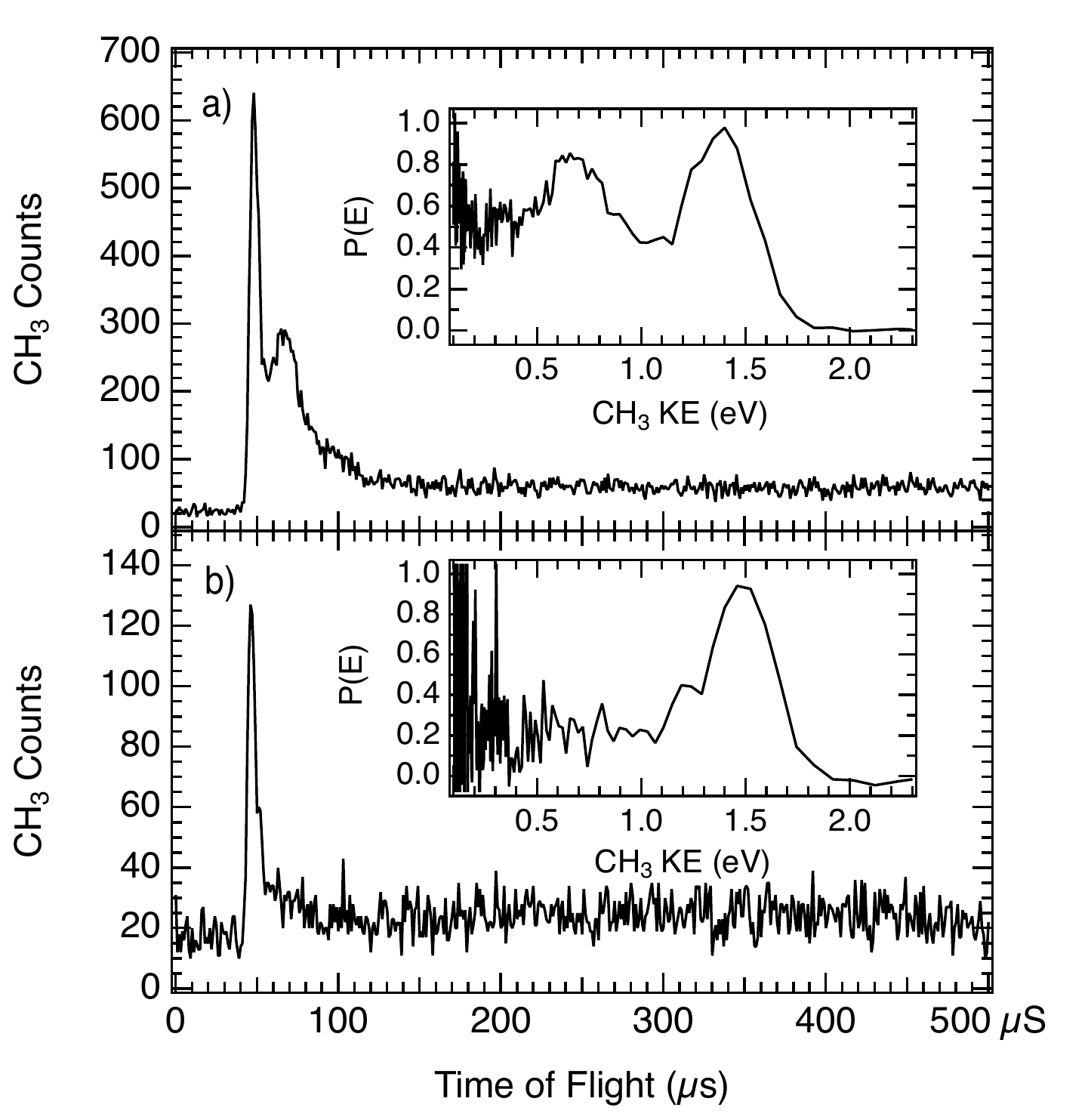}
\caption{\label{Fig_CH3Cl_193nm} Time of flight spectra for CH\(_3\) photofragments due to the photodissociation of 1ML CH\(_3\)Cl on 3ML (a) C\(_6\)H\(_6\) and (b) C\(_6\)F\(_6\) thin films, obtained using 193nm light. The inset plots show the corresponding CH$_3$ kinetic energy distributions. The CH$_3$ photofragments are detected in the surface normal direction. This data was obtained on a Cu(100) substrate.}
\end{figure}

The 193nm photodissociation of 1ML CH\(_3\)Cl adsorbed on 3ML of C\(_6\)H\(_6\) results in a TOF spectrum for CH\(_3\) photofragments displaying two peaks as seen in Fig.~{\ref{Fig_CH3Cl_193nm}}(a)-- a fast peak at 48\(\mu\)s flight time (1.40eV) and a slower peak at 66\(\mu\)s (0.65eV). The fast peak is consistent with that expected for neutral photodissociation, and the slower peak consistent with photoelectron mediated CT-DEA. These observed features are broadly consistent with surface photodissociation of CH$_3$Cl seen in other heterogeneous systems such as CH$_3$Cl/D$_2$O/Cu(110)\cite{Jensen:2015dg}. The magnitude of these features correlate with the CH$_3$Cl coverage for a fixed amount of C$_6$H$_6$. Photodissociation of similarly prepared 1ML CH\(_3\)Cl on 3ML C\(_6\)F\(_6\) results in a TOF spectrum of Fig.~{\ref{Fig_CH3Cl_193nm}}(b) that shows the fast neutral photodissociation feature but the CT-DEA feature is entirely absent. Other differences between the CH\(_3\) TOF spectra on the two C\(_6\)X\(_6\) thin films are that the signal is significantly weaker for CH$_3$Cl/C\(_6\)F\(_6\) and that the peak in the TOF spectra is consistently found to be 1--2\(\mu\)s faster than for CH$_3$Cl/C$_6$H$_6$, seen in both the leading edge onset and the peak center, shifting the centre of the P(E) distribution $\sim$0.10eV higher on C$_6$F$_6$. This could be due to altered dynamics (Equ. {\ref{Equ_3}}), for example if the Cl partner is less free to move toward the surface during dissociation. The significantly increased yield and the presence of the CT-DEA dissociation pathway observed for the CH$_3$Cl/C$_6$H$_6$ thin films in Fig.~{\ref{Fig_CH3Cl_193nm}}a correlate with the CH$_3$Cl depletion cross sections observed for these systems at 193nm. On the C$_6$H$_6$ thin film the cross section was found to be $2.8\times10^{-19}$cm$^2$ as compared to $1.0\times 10^{-19}$cm$^2$ on the C$_6$F$_6$ thin film. The corresponding gas-phase photodissociation cross section for  CH$_3$Cl is $0.70\times10^{-19}$cm$^2$, at 193nm\cite{KellerRudek:2013wf}.

\begin{figure}
\includegraphics[scale=0.55]{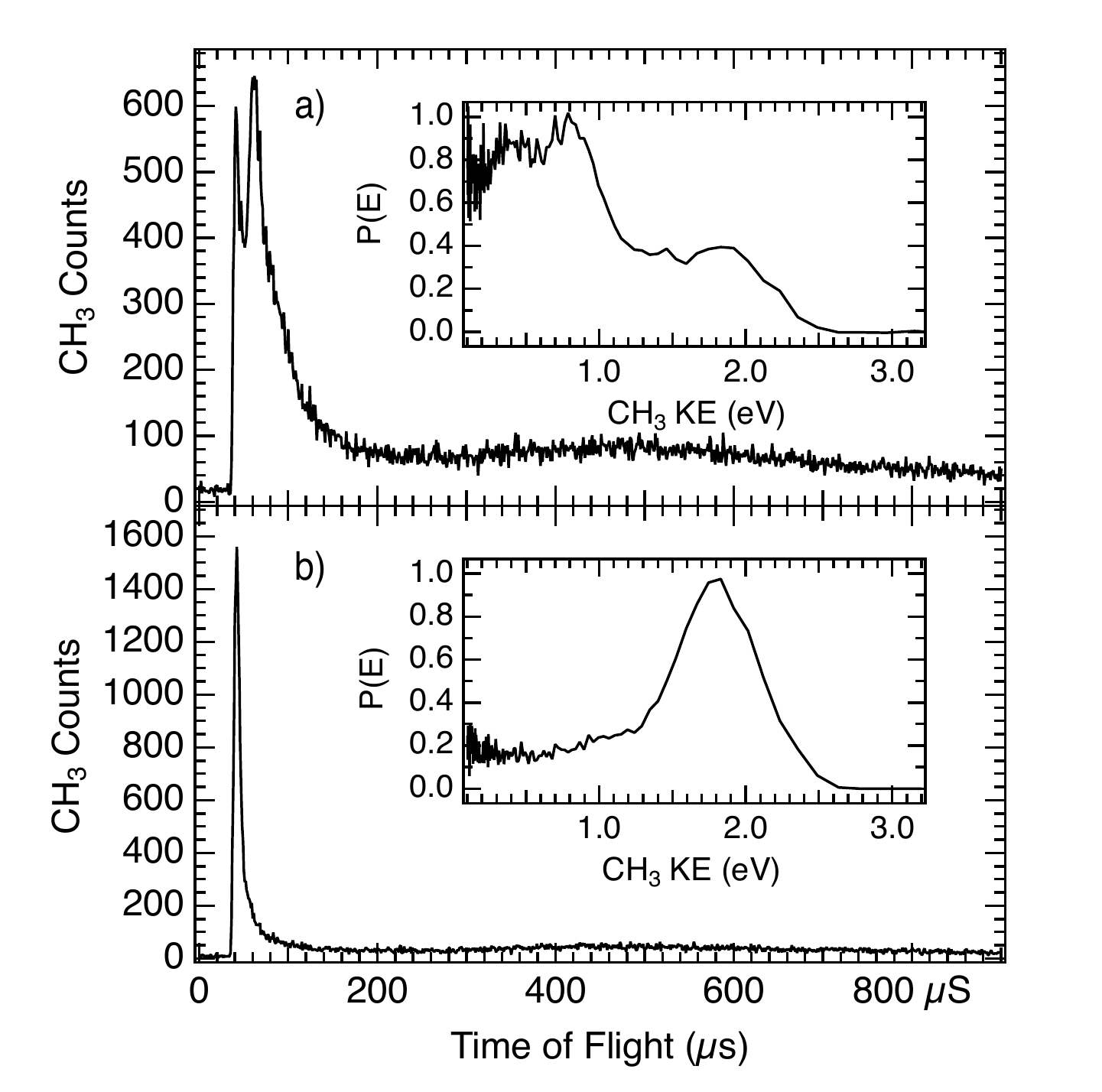}
\caption{\label{Fig_CH3Br_193nm} Time of flight spectra for CH\(_3\) photofragments due to the photodissociation of 1ML CH\(_3\)Br on 3ML (a) C\(_6\)H\(_6\) and (b) C\(_6\)F\(_6\) thin films using 193nm light. The inset plots show the corresponding CH$_3$ kinetic energy distributions. This data was obtained on a Cu(100) substrate.}
\end{figure}

The photodissociation of CH$_3$Br on 3ML C$_6$X$_6$ thin films using 193nm light (Fig.~{\ref{Fig_CH3Br_193nm}}) finds that CH$_3$Br/C$_6$H$_6$ has a CH$_3$ photofragment TOF spectrum with both a neutral photodissociation (peak at \(\sim\)41\(\mu\)s, 1.9eV) and photoelectron CT-DEA (peak at 62\(\mu\)s, 0.8eV) features. These peak assignments and energies are similar to those seen previously for CH$_3$Br/D$_2$O/Cu(110)\cite{Jensen:2015dg}. For CH\(_3\)Br adsorbed on a 3ML C\(_6\)F\(_6\) thin film, the TOF spectra display only the neutral photodissociation peak. In contrast to the case for CH$_3$Cl/C$_6$F$_6$, the CH$_3$Br/C$_6$F$_6$ has an {\it{enhanced}} CH$_3$ signal in the TOF spectra as compared to CH$_3$Br/C$_6$H$_6$, and the neutral photodissociation TOF peak occurs at the same flight time. The depletion cross sections for adsorbed CH$_3$Br at 193nm are found to be $2.9\times 10^{-18}$cm$^2$ when adsorbed on 3ML C$_6$H$_6$ thin films as compared to $9.9\times 10^{-19}$cm$^2$ on a 3ML C$_6$F$_6$ thin film. These cross sections can be compared to the gas-phase value of $5.6\times10^{-19}$cm$^2$ for CH$_3$Br at 193nm\cite{KellerRudek:2013wf}.

The angular distributions we have measured for several of these thin film systems (including those for CH$_3$I discussed below) find that the CH$_3$ photofragment yield is peaked in the surface normal direction, and diminishes $\propto cos^N(\theta)$ where $N\sim 6-8$ ($\theta$ is the angle measured from normal). This indicates that these methyl halides on the C$_6$X$_6$ thin films are generally organized such that at least a substantial proportion of the molecules have the CH$_3$ moiety pointed in the surface normal direction. This would be consistent with an antiferroelectric structure for these dipolar molecules, which have been proposed for methyl halides in other studies for a variety of substrates\cite{Rowntree:1990cm,Fairbrother:1994vj,Nalezinski:1997uo} where ordering is dominated by the molecular dipole moment rather than chemical interactions between the halogen moiety and a metal surface. We are not aware of any previous studies of structural ordering for methyl halides adsorbed on the C$_6$X$_6$ thin films. We do see some evidence for a different orientation at lower methyl halide coverages, which is one reason that the present study focussed on monolayer coverages for the methyl halides.

\begin{figure}
\includegraphics[scale=0.60]{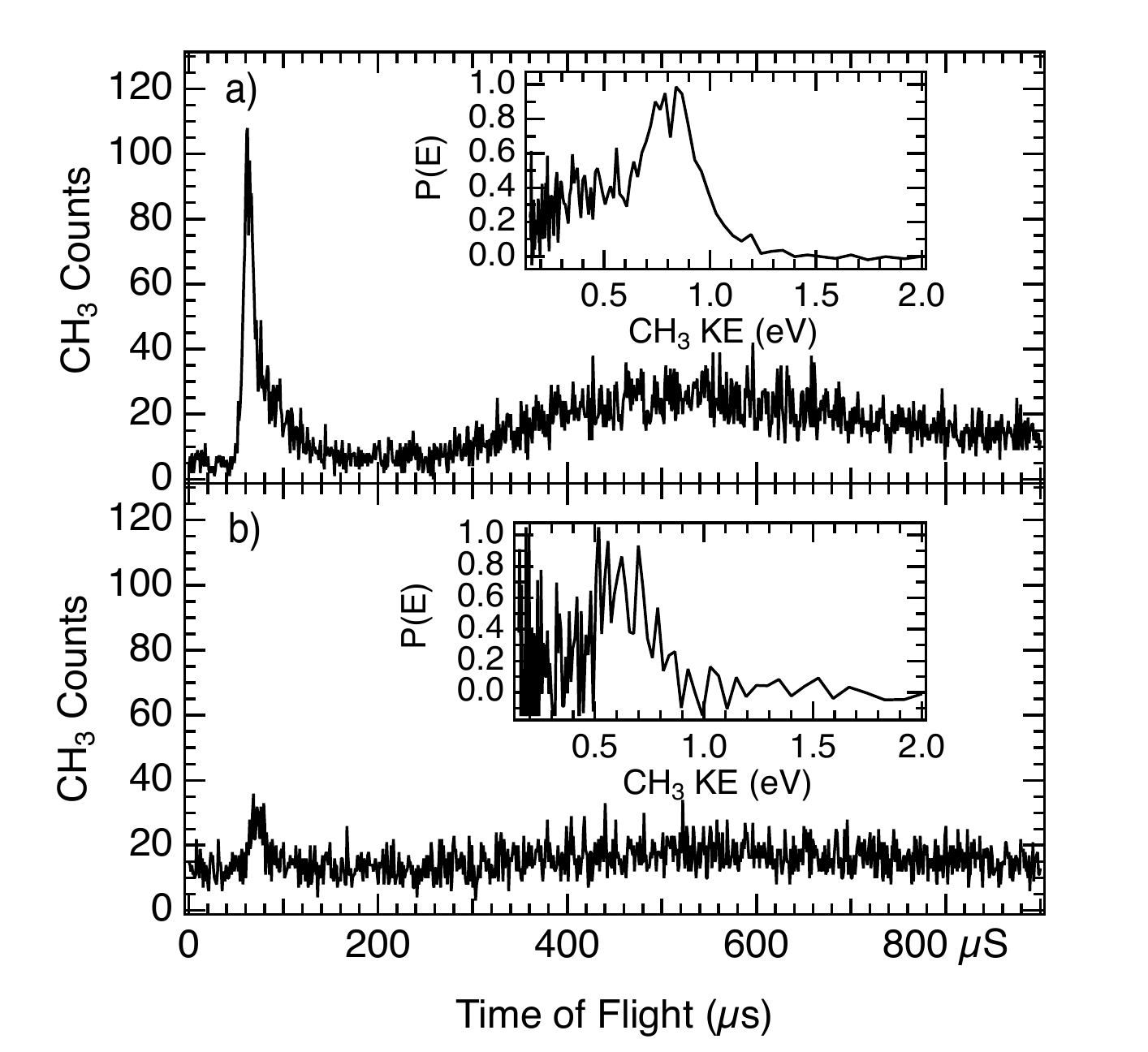}
\caption{\label{Fig_CH3X_248nm} TOF spectra for CH\(_3\) photofragments from a monolayer of (a) CH\(_3\)Br and (b) CH\(_3\)Cl on a 3ML C\(_6\)H\(_6\) thin film on Cu(110), using 248nm light. The inset plots show the corresponding CH$_3$ kinetic energy distributions.}
\end{figure}

Photodissociation of CH$_3$Cl or CH$_3$Br adsorbed on a 3ML C$_6$H$_6$ thin films using longer wavelength 248nm light finds that these methyl halides are photodissociated, albeit with much lower CH$_3$ photofragment yields, as shown in Fig.~{\ref{Fig_CH3X_248nm}}. The CH$_3$ photofragment TOF features observed using 248nm light for both methyl halides are consistent with CT-DEA, with peaks at the same CH$_3$ photofragment flight times and kinetic energies (0.8eV for CH$_3$Br; 0.6eV for CH$_3$Cl) as for the corresponding 193nm photodissociation data. The neutral photodissociation features seen for these methyl halides using 193nm light are absent for 248nm light, which is consistent with the gas-phase photodissociation cross sections for both methyl halides being at least 2 orders of magnitude smaller at this wavelength as compared to 193nm\cite{KellerRudek:2013wf}. The depletion cross section at 248nm for CH$_3$Br/C$_6$H$_6$ thin film is measured to be $3.0\times 10^{-19}$cm$^2$, reduced by roughly a factor of 10 compared to the same system at 193nm. The cross section for CH$_3$Cl on this thin-film was so low that we could not reliably measure it. When these methyl halides are adsorbed on C$_6$F$_6$ thin films, no CH$_3$ photofragment TOF signals were observed using 248nm light. This is consistent with the observations from Figs.~{\ref{Fig_CH3Cl_193nm}} -- {\ref{Fig_CH3X_248nm}} in that the CT-DEA pathway is not observed for these methyl halides on the C$_6$F$_6$ thin films and that the neutral photodissociation has a cross section too low to be observable. 

\begin{figure}
\includegraphics[scale=0.65]{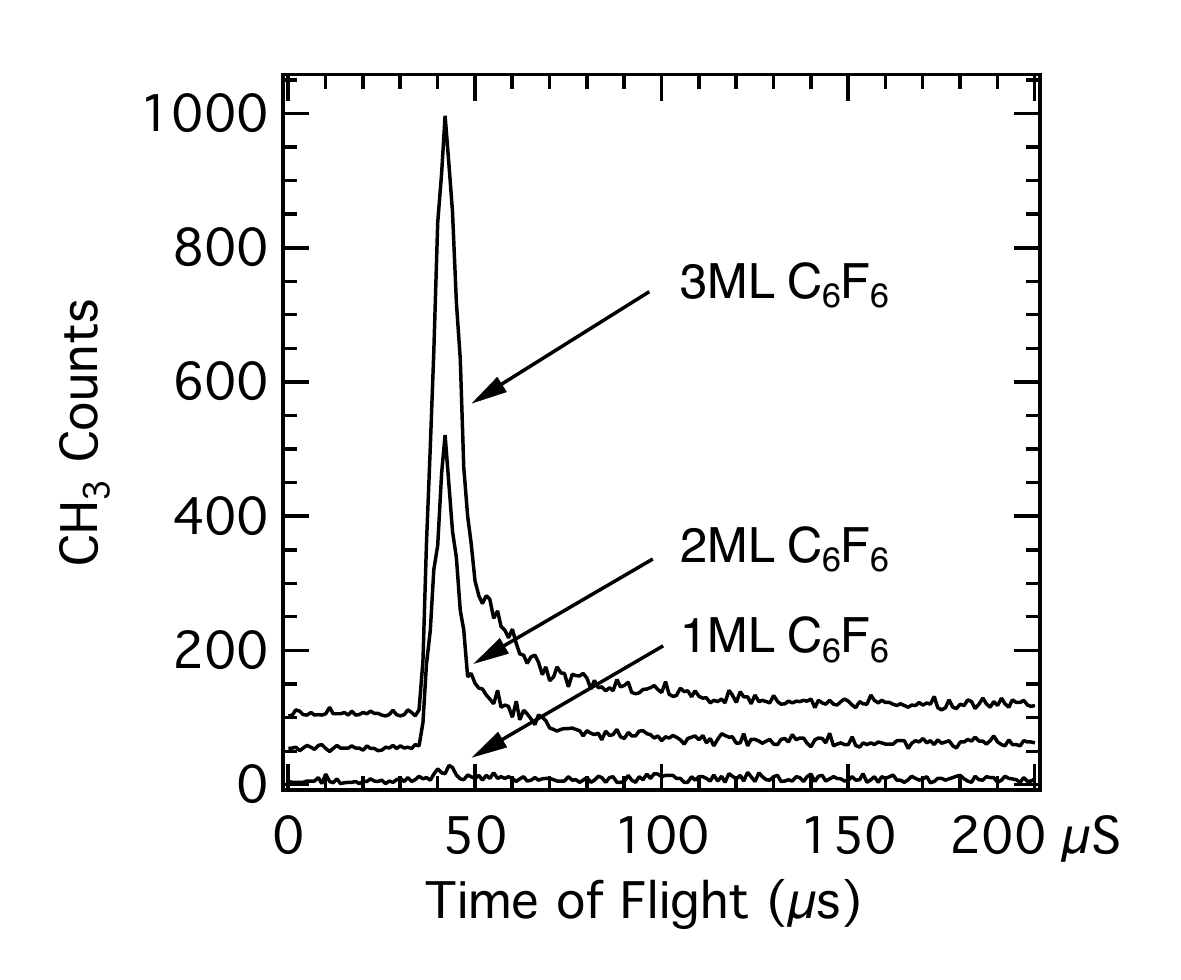}
\caption{\label{Fig_CH3Br_vary_C6F6} TOF spectra for CH\(_3\) photofragments from photodissociation of 1ML CH\(_3\)Br (\(\lambda\)=193nm) on varied thin films of C\(_6\)F\(_6\) on Cu(110). The spectra are vertically separated by 50 counts for clarity.}
\end{figure}

\begin{figure}
\includegraphics[scale=0.65]{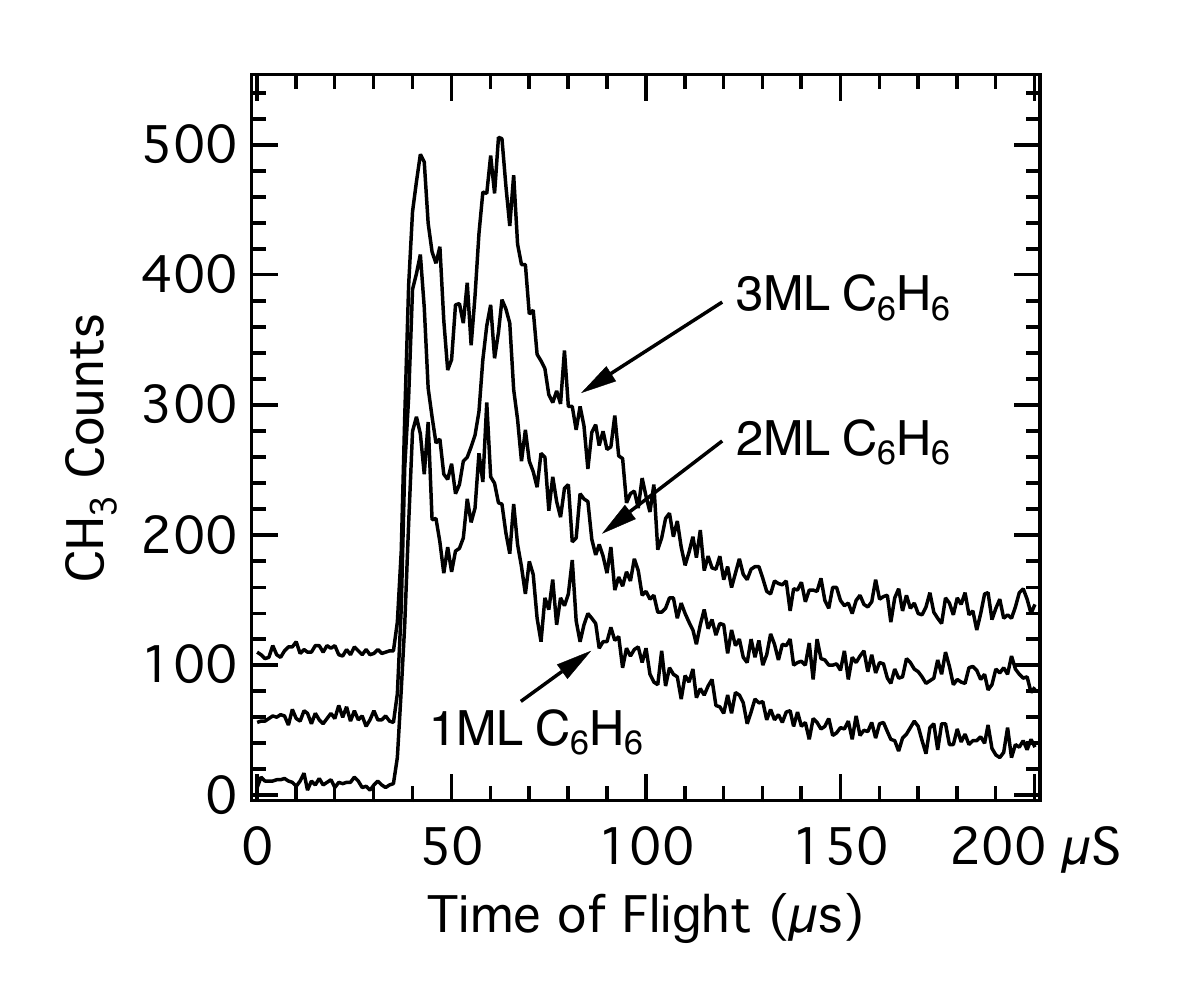}
\caption{\label{Fig_CH3Br_vary_C6H6} TOF spectra for CH\(_3\) photofragments from photodissociation of 1ML CH\(_3\)Br (\(\lambda\)=193nm) on varied  thin films of C\(_6\)H\(_6\) on Cu(110). The spectra are vertically separated by 50 counts for clarity.}
\end{figure}

A sequence of TOF spectra from 193nm photodissociation of CH$_3$Br/C$_6$F$_6$ thin films of varying thickness are shown in Fig.~{\ref{Fig_CH3Br_vary_C6F6}}. When the CH$_3$Br is adsorbed on a 1ML C$_6$F$_6$ film, essentially no CH$_3$Br photodissociation is observed in the TOF spectra. However when the film thickness is increased to 2ML and higher, a clear neutral photodissociation signal is observed. That the neutral photodissociation pathway is absent on the 1ML C$_6$F$_6$ thin film is also observed for CH$_3$Cl and CH$_3$I photodissociation, so appears to be a common feature for the halomethanes in this study. This finding is discussed further in Section \ref{no_dea_in_c6f6}. The 193nm photodissociation of CH$_3$Br on varied thin films of C$_6$H$_6$ is shown in Fig.~{\ref{Fig_CH3Br_vary_C6H6}}. In this case, both the neutral as well as CT-DEA photodissociation pathways are observed at all C$_6$H$_6$ film thicknesses including for the 1ML C$_6$H$_6$ thin film. This behaviour is also seen for the CH$_3$Cl/C$_6$H$_6$ system. 

\subsection{Photodissociation of \texorpdfstring{CH$_3$I}{CH3I} on \texorpdfstring{C$_6$F$_6$}{C6F6} Thin Films}
\begin{figure}
\includegraphics[scale=0.57]{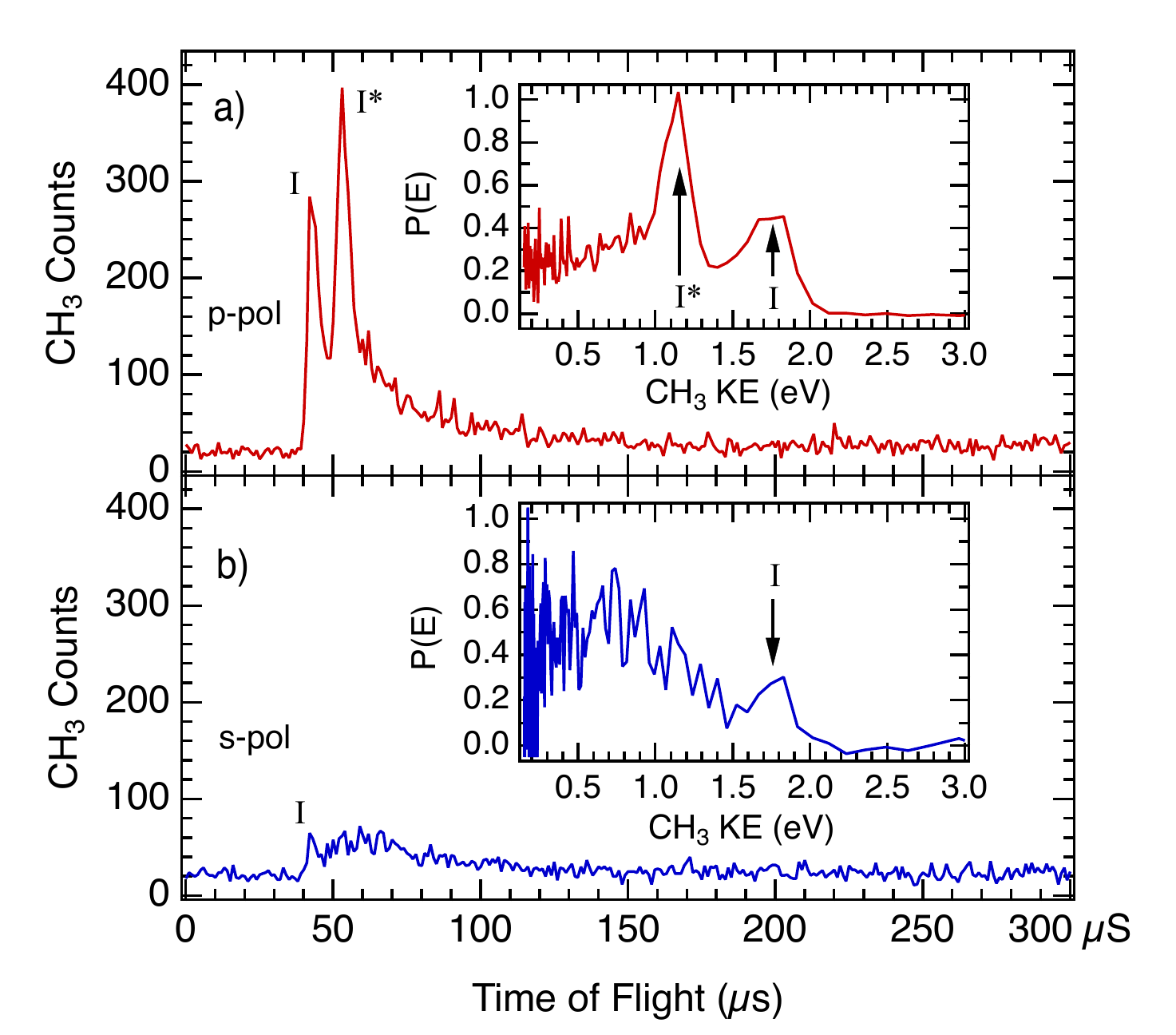}
\caption{\label{Fig_CH3I_C6F6_p_s} TOF spectra for CH$_3$ photofragments from the \(\lambda\)=248nm photodissociation of 1ML CH$_3$I on a 3ML C$_6$F$_6$ thin film on Cu(100) using (a) p-polarized light and (b) s-polarized light. The features at 42$\mu$s flight time (1.75eV) are labelled ``I'' and that at 53$\mu$s (1.15eV) is labelled ``I$^*$'', as discussed in the text. }
\end{figure}

Following on from the observations made for CH$_3$Cl and CH$_3$Br on thin C$_6$X$_6$ films, we have also investigated the photodissociation of CH$_3$I on these thin films. In Fig.~{\ref{Fig_CH3I_C6F6_p_s}} the CH$_3$ photofragments detected in the surface normal direction from 248nm photodissociation of 1ML CH$_3$I on a 3ML C$_6$F$_6$ thin film is shown for p- and s-polarized light incident at 45$^\circ$ from the surface normal. For the p-polarized incident light (Fig.~{\ref{Fig_CH3I_C6F6_p_s}}a), the neutral photodissociation results in two peaks in the TOF spectrum similar to that seen in several other surface photodissociation studies\cite{Jensen:2005js}. The two prominent peaks in the TOF spectra are the result of neutral photodissociation along two pathways as per Equ.~{\ref{Equ_1}}, the faster one (42$\mu s$) for the outcome CH$_3$ + I($^2P_{3/2}$), and the slower CH$_3$ peak (53$\mu s$) for the pathway leading to the I*($^2P_{1/2}$) as the outcome. When s-polarized light is incident (Fig.~{\ref{Fig_CH3I_C6F6_p_s}}b), a much lower CH$_3$ photofragment signal is observed in the surface normal direction. The pronounced differences in the photofragment yields from the p- and s-polarized light are a consequence of the selection rule for the $X-{^3Q_0}$ transition in the A-band leading to this dissociation-- requiring an $\vec{E}$-field component parallel the C--I molecular axis\cite{Eppink:1998ue}. For the rapid bond scission in this dissociation process, the CH$_3$ photofragments will be observed if they are ejected in the direction of the detector (in Fig.~{\ref{Fig_CH3I_C6F6_p_s}}, the surface normal). Given the geometry of this experiment, this requirement is met for p-polarized light while it is not met for s-polarized light. In addition, it is likely that the CH$_3$I molecular orientation on the C$_6$F$_6$ thin film plays a significant role. The angular distributions of the CH$_3$ photofragment yields we have measured for this system are peaked in the surface normal direction,  diminishing proportional to $cos^N(\theta)$ where $N\sim 7-8$. These observations suggest that a substantial fraction of the adsorbed CH$_3$I are aligned with the CH$_3$ moiety pointed at or close to the surface normal direction (e.g. the antiferroelectric structure as discussed in Section {\ref{CH3X_results}}), so that the CH$_3$ photofragments leave the surface region without substantial inelastic interactions during or following dissociation. The s-polarized light can dissociate CH$_3$I molecules that have the molecular axis lying more parallel to the substrate. If this orientation were present, the CH$_3$ photofragments would not reach the detector directly. It is possible that such a photofragment could scatter such that they ultimately travel in the direction of the surface normal, and then be detected in Fig.~{\ref{Fig_CH3I_C6F6_p_s}}b. This might be the process responsible for the low levels of CH$_3$ photofragments that are being detected using s-polarized light-- we note that there is a higher proportion of the detected CH$_3$ photofragments that have lower translational energy (inset of Fig.~{\ref{Fig_CH3I_C6F6_p_s}}b). Alternatively, it is possible that upright CH$_3$I molecules could be photodissociated via the weaker $X-{^1Q_1}$ transition in the A-band, which is perpendicular\cite{Eppink:1998ue, Jensen:2005js}. This transition almost exclusively leads to the outcome pathway of CH$_3$ + I in the gas-phase\cite{Townsend:2004uu}, so is consistent with the observed small peak at 42$\mu s$ while the slower I* peak is absent, as seen in Fig.~{\ref{Fig_CH3I_C6F6_p_s}}b. The depletion cross section observed for 1ML CH$_3$I on 3ML C$_6$F$_6$/Cu was found to be $8.2\times10^{-19}$cm$^2$, which can be compared to the gas-phase value\cite{KellerRudek:2013wf} of $8.5\times10^{-19}$cm$^2$. 

\subsection{Photodissociation of \texorpdfstring{CH$_3$I}{CH3I} on \texorpdfstring{C$_6$H$_6$}{C6H6} Thin Films}
\begin{figure}
\includegraphics[scale=0.60]{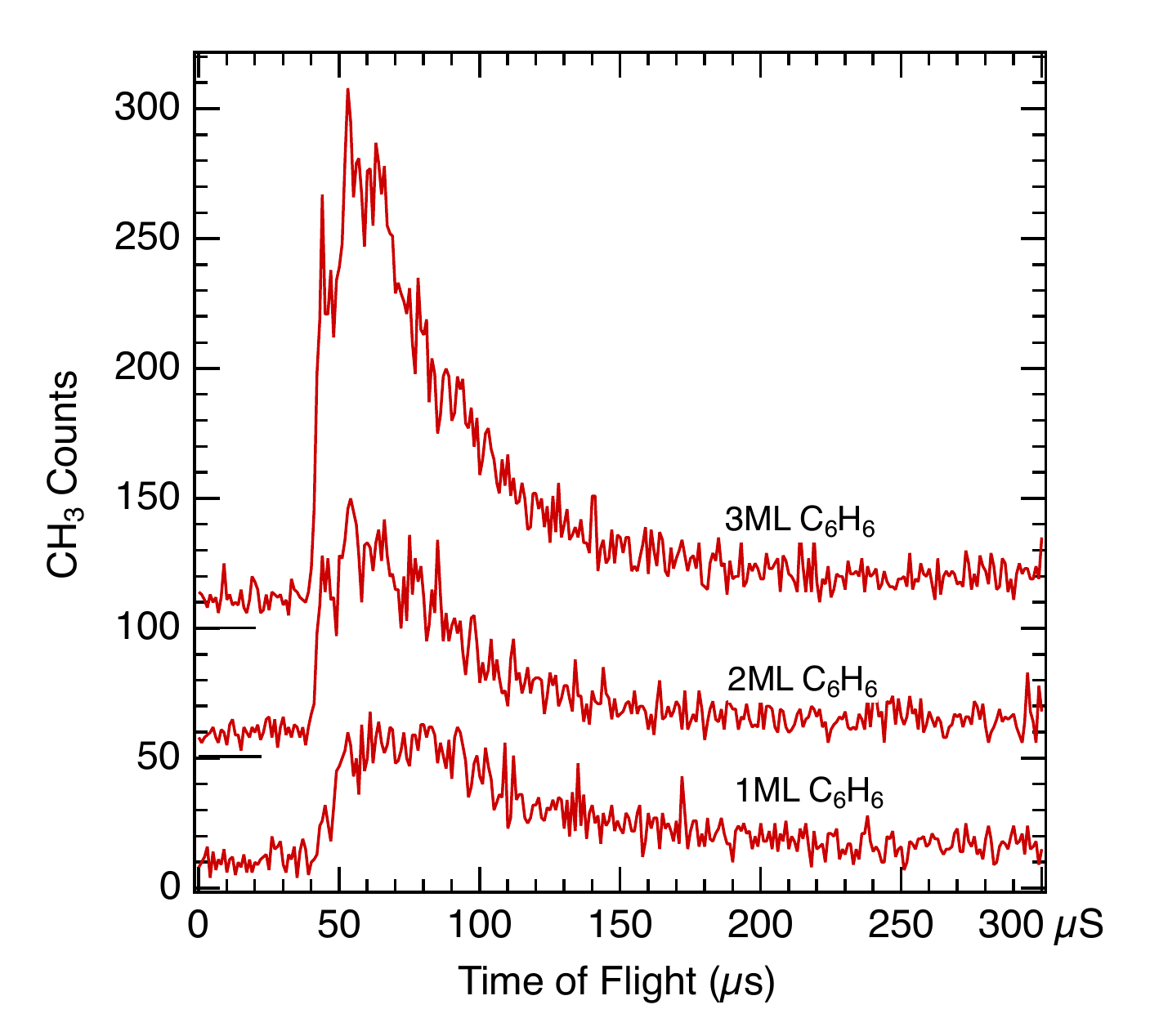}
\caption{\label{Fig_CH3I_vary_C6H6_p} TOF spectra for 1ML CH$_3$I on C$_6$H$_6$ thin films (1--3ML) on Cu(100). These spectra were obtained using 248nm p-polarized light. The spectra are vertically separated by 50 counts for clarity.}
\end{figure}

\begin{figure}
\includegraphics[scale=0.60]{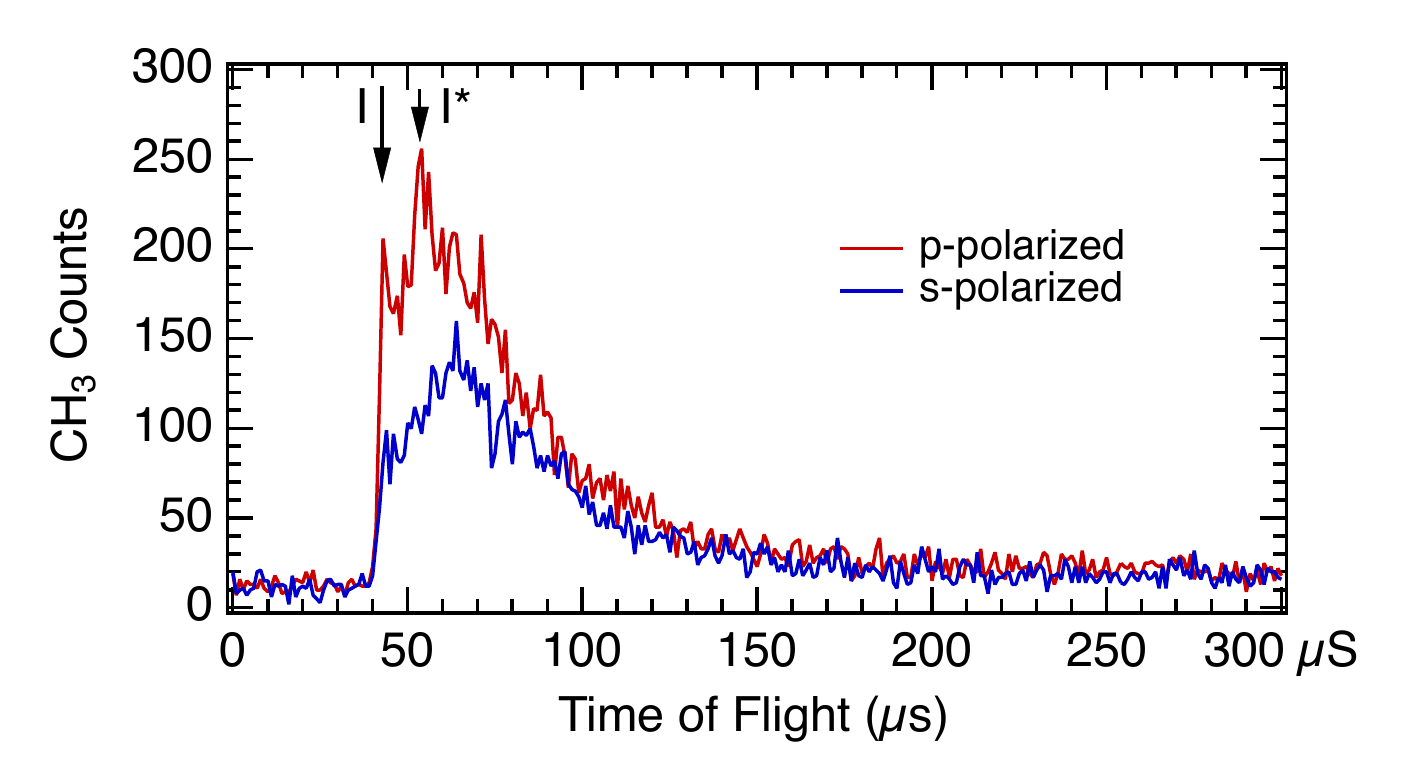}
\caption{\label{Fig_CH3I_C6H6_p_s} TOF spectra from 1ML CH$_3$I on a 4ML C$_6$H$_6$ thin film on Cu(100), using p- and s-polarized 248nm light. The peaks in the spectrum obtained using p-polarized light are found at 43$\mu s$ and 54$\mu s$, consistent with the peaks labelled $I$ and $I^*$  in Fig.~{\ref{Fig_CH3I_C6F6_p_s}}a.}
\end{figure}

Time-of-flight spectra from 1ML CH$_3$I adsorbed on varying thin films of C$_6$H$_6$ using p-polarized 248nm light are shown in Fig.~{\ref{Fig_CH3I_vary_C6H6_p}}. Similar to the observations for CH$_3$Cl and CH$_3$Br on C$_6$H$_6$ thin films (e.g. Fig.~{\ref{Fig_CH3Br_vary_C6H6}}), CH$_3$ photofragments are observed from C$_6$H$_6$ thin films of 1ML and higher thicknesses. In contrast to the other methyl halides, the TOF spectra from CH$_3$I are rather broad and though some individual peaks are visible, the delineation of them is less clear, and there is no obvious component that is a consequence of the CT-DEA mechanism. The TOF spectra in Fig.~{\ref{Fig_CH3I_C6H6_p_s}} show the CH$_3$ photofragments from 1ML CH$_3$I on 4ML C$_6$H$_6$, where the incident light is switched between p- and s-polarized light. While there are clear differences in the spectra for the two polarizations, the ``switching'' effect is not as pronounced as was the case for the CH$_3$I on C$_6$F$_6$ thin films (Fig.~{\ref{Fig_CH3I_C6F6_p_s}}), where the I and I$^*$ pathways can be seen in the TOF spectrum peaks for p-polarized light, and nearly absent for s-polarized light. In trying to understand the CH$_3$ photofragment TOF features for this system, which could be due to a variety of possible mechanisms (e.g. CT-DEA, mixed C-I orientations), we have found that the clearest understanding comes from examining TOF spectra obtained using somewhat thicker C$_6$H$_6$ films

\begin{figure}
\includegraphics[scale=0.60]{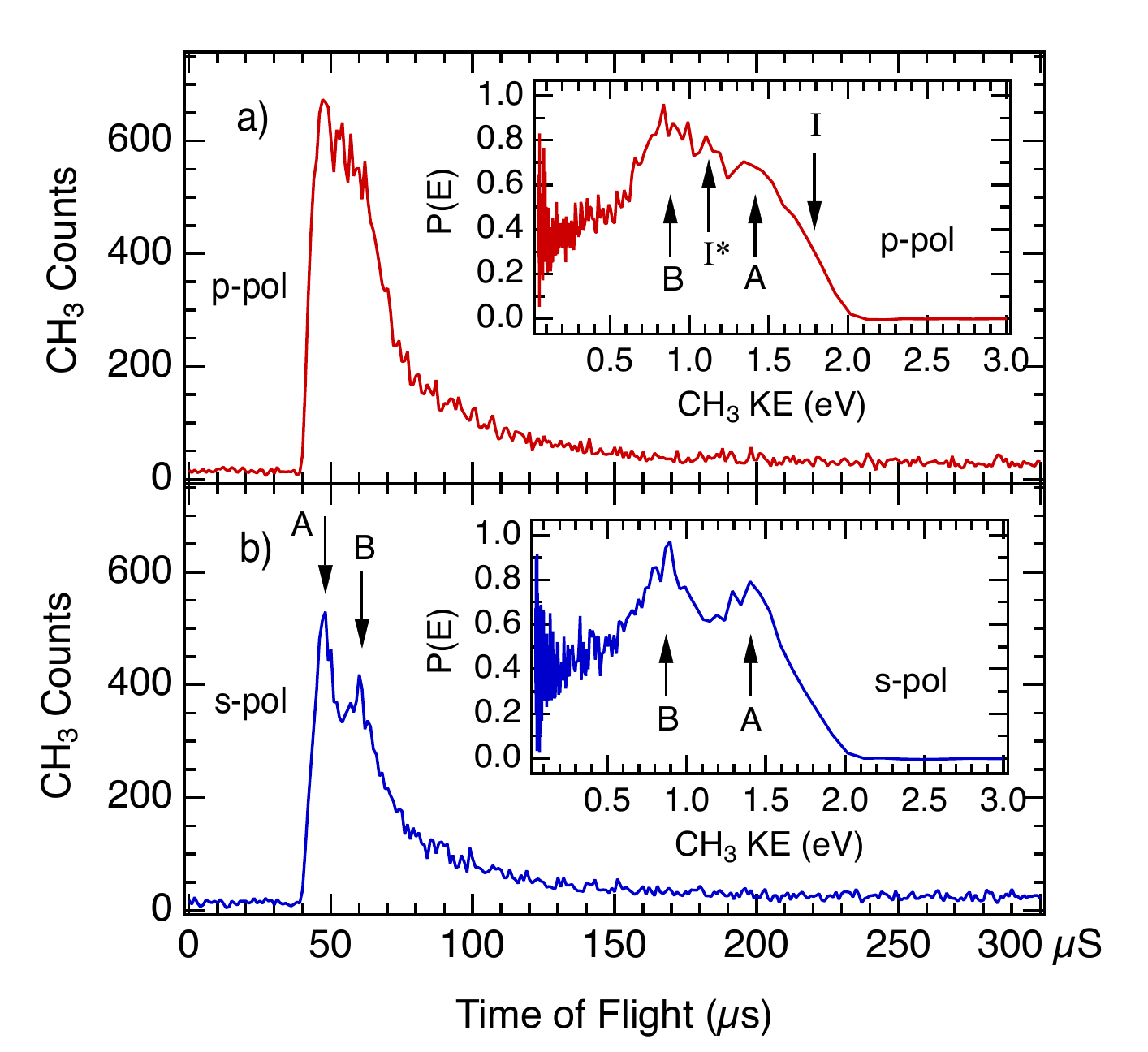}
\caption{\label{Fig_CH3I_thick_C6H6_p_s} Photodissociation TOF spectra for 1ML CH$_3$I on a 10ML C$_6$H$_6$ thin film on Cu(100), obtained using (a) p-polarized and (b) s-polarized  248nm light. The features labelled "A" and "B" are observed at 47$\mu$s (1.4eV) and 60$\mu$s (0.9eV) respectively. The features labelled I and I* correspond to those identified in Fig.~{\ref{Fig_CH3I_C6F6_p_s}}. }
\end{figure}

The TOF spectra from photodissociation of 1ML CH$_3$I on 10ML C$_6$H$_6$ are shown in Fig.~{\ref{Fig_CH3I_thick_C6H6_p_s}}, obtained using p- and s-polarized 248nm light. The TOF spectrum obtained using s-polarized light (Fig.~{\ref{Fig_CH3I_thick_C6H6_p_s}}b) displays distinct peaks at 47$\mu$s (labelled {\it{``A''}}) and 60$\mu$s (labelled {\it{``B''}}) flight times, and the corresponding CH$_3$ fragment kinetic energy distributions locates these peaks at 1.4eV and 0.9eV respectively. These peaks are distinct from those seen for CH$_3$I/C$_6$F$_6$ in Fig.~{\ref{Fig_CH3I_C6F6_p_s}}a and the spectra are also very different from Fig.~{\ref{Fig_CH3I_C6F6_p_s}}b that also used s-polarized light. These peaks A and B are most clear when the spectra are obtained using thicker C$_6$H$_6$ films but are apparent in TOF spectra obtained for thinner films, particularly when using s-polarized light. When p-polarized light is used (Fig.~{\ref{Fig_CH3I_thick_C6H6_p_s}}a), the TOF spectrum has overall higher signal, but individual peak features are overlapping, so hard to discern. In the corresponding CH$_3$ photofragment energy distribution inset, one can identify 4 features-- the same peaks A and B as seen in the s-polarized data, as well as peaks due to the I and I* channels such as those observed in Fig.~{\ref{Fig_CH3I_C6F6_p_s}}a. Though these features are overlapping, these peaks can be identified in similar data over a range of C$_6$H$_6$ film thicknesses through the differences between spectra obtained using p- and s-polarized light. For thinner C$_6$H$_6$ films, the photodissociation peaks associated with the I and I* pathways can be seen more clearly in spectra using p-polarized light (e.g. Figs.~{\ref{Fig_CH3I_vary_C6H6_p}} and {\ref{Fig_CH3I_C6H6_p_s}}) while the A and B features are less well developed in the thinner C$_6$H$_6$ films. We investigated the photodissociation of CH$_3$I on thicker (e.g. 20ML) C$_6$H$_6$ films but did not find clearer signals than seen for the 10ML C$_6$H$_6$ films. The additional pathway for photodissociation is reflected in an increased depletion cross-section-- for 1ML CH$_3$I on 2ML C$_6$H$_6$/Cu a cross section of $9.3\times10^{-19}$cm$^2$ was found. The spectra of Figs.~\ref{Fig_CH3I_vary_C6H6_p}--{\ref{Fig_CH3I_thick_C6H6_p_s}} suggest that in addition to neutral photodissociation similar to that of gas-phase CH$_3$I, for CH$_3$I on C$_6$H$_6$ there is a new photodissociation pathway operative at 248nm, and the available evidence suggests that this is due to initial photoabsorption involving the C$_6$H$_6$ adlayer. This pathway is relatively insensitive to the laser light polarization and involves heteromolecular energy transfer following photoabsorption, resulting in the CH$_3$I dissociation features labelled A and B in the TOF spectra. This possibility is discussed further in Sect. \ref{sect_ch3I_c6h6}.

\section{Additional Discussion}
\label{additional_discussion}
\subsection{Photodissociation of \texorpdfstring{CH$_3$I}{CH3I} on \texorpdfstring{C$_6$H$_6$}{C6H6}}
\label{sect_ch3I_c6h6}
The observation of the features A and B in the $\lambda$=248nm photodissociation of CH$_3$I on C$_6$H$_6$ thin films requires a discussion of what possible mechanisms are responsible for these. It has been longstanding for surface photodissociation studies on metal surfaces that photoelectrons and/or hot photoelectrons can lead to bond scission through dissociative electron attachment (DEA). We do not believe that such a process is responsible for the A and B features observed in Fig.~{\ref{Fig_CH3I_thick_C6H6_p_s}}. Dissociative electron attachment by low-energy electrons for CH$_3$I in condensed environments has been found to be suppressed\cite{Jensen:2008jv,Fabrikant:2011ep} due to a shielding of the long-range electron molecule interaction, in contrast to the enhanced DEA seen for other halomethanes and related molecules when adsorbed. Indeed, in previous surface photodissociation studies where DEA of adsorbed CH$_3$I is anticipated, it is a small secondary channel\cite{Miller:2013eq} and is seen at longer flight times (broader peak near 70--80$\mu$s or 0.50eV). We also compared the photodissociation of CH$_3$I on multilayer C$_6$H$_6$ films when it was adsorbed on Cu(100) to the same adsorption system on Cu(100)-Cl, and no differences were seen in the TOF spectra. In other adsorption systems studied using 248nm light where photoelectron CT-DEA is understood to be the mechanism of dissociation, we have found that such photodissociation is diminished or eliminated by chlorinating the Cu surface.\footnote{The details of the impact for photoelectron generation between the Cu(100) and Cu(100)-Cl substrates are unclear, as the net changes in workfunction from the combined adsorbates are not known. Ref.~\onlinecite{Goddard:1977} found $\Delta\Phi$=+1.2eV for Cu(111)-Cl and Ref.~\onlinecite{Sonoda:2007uz} found $\Delta\Phi$=-0.66eV for 2ML C$_6$H$_6$ on Cu(110). We note that the 248nm (5.0eV) photons used are close in energy to the workfunction of clean Cu(100) and that previous work in our lab has found substantially reduced CT-DEA related halomethane photodissociation on the chlorinated surface.}  Finally, the illumination of CH$_3$Br on thick C$_6$H$_6$ films using $\lambda$=248nm found that the CT-DEA features (e.g. Fig. {\ref{Fig_CH3X_248nm}a}) diminished to undetectable levels, which is similar to the observations for the same methyl halide on D$_2$O and CH$_3$OH films in which the CT-DEA yield also diminished for thicker films\cite{Jensen:2015dg}. We would expect similar behaviour for CH$_3$I if CT-DEA from substrate photoelectrons was the operative mechanism for either or both of the A and B features observed on C$_6$H$_6$ films.

That the photodissociation features A and B in Fig.~{\ref{Fig_CH3I_thick_C6H6_p_s}} could be the result of photon absorption in the molecular thin film is supported by the following. Gas-phase C$_6$H$_6$ has an absorption band ${^1B_{2u}}\leftarrow {^1A_{1g}}$ in the near UV (230--265nm) region, with  cross sections\cite{KellerRudek:2013wf,Dawes:2017dpa} of the order of $2\times10^{-18}cm^2$.  In the gas-phase, the excited singlet state is non-dissociative, with a fluorescence lifetime of tens of nanoseconds\cite{Kovacs:2009ko}. This absorption band is also observed in condensed C$_6$H$_6$, with the detailed vibronic structure shifted depending on the crystalline structure\cite{Dawes:2017dpa}. For monolayer C$_6$H$_6$ films on Cu, a UV Reflection-Absorption Spectroscopy study observed the vibronic structure in spectra at these wavelengths as well\cite{Peng:2000up}. Photoexcitation of benzene by UV photons in this range has been found to be associated with photodesorption of C$_6$H$_6$ from thick films, and was sensitive to the tuning of the UV wavelength to align with specific absorption lines in the ${^1B_{2u}}\leftarrow {^1A_{1g}}$ band\cite{Thrower:2008tx}. Of relevance for the KrF laser light used in this work, the long-wavelength tail of the 248nm excimer laser bandwidth overlaps vibronic bands observed for condensed phase C$_6$H$_6$ (the peak labelled $A_2$ in Ref.~\onlinecite{Peng:2000up} at 248.7nm; the vibronic peak $6^1_01^2_0$ for condensed C$_6$H$_6$ at 90K in Ref.~\onlinecite{Dawes:2017dpa} at 248.85nm; at 248.7nm in Ref.~\onlinecite{Stubbing:2020fy}). That near-UV photon absorption by the CH$_3$I--C$_6$H$_6$ combination could result in CH$_3$I dissociation is suggested by the work of Dubois and Noyes\cite{Dubois:1951ub}, in which gas-phase mixtures of C$_6$H$_6$ and CH$_3$I were found to lead to enhanced CH$_3$I dissociation (observed as formation of C$_2$H$_6$(g)) when illuminated by a mercury lamp (253.7nm). Notably in that work, mixtures of CH$_3$Cl and CH$_3$Br with C$_6$H$_6$ showed no discernable enhancement when similarly illuminated. Quenching of near-UV photoexcited C$_6$H$_6$ by halogenated species has also been studied in the gas-phase, with large cross section for heavier halogen-containing molecules\cite{Gupta:1972cc}. It should also be noted that benzene complexes with halogenated molecules are well known to have modified electronic and energetic structure\cite{Zwier:1991gn}, which can be significant for heavy atoms such as Iodine, which can also promote singlet-triplet intersystem crossings. More recent work studied charge-transfer complexes for C$_6$H$_6$--I$_2$ in clusters and in liquids, examining the photodissociation pathways in the near-UV for in some detail\cite{DeBoer:1996df,Cheng:1995gr}. 

If it is assumed that the features A and B are the result of intermolecular energy transfer, the initial energy required (equivalent to $h\nu$ in Equ. {\ref{Equ_3}}) can be estimated, assuming a similar partitioning of energy seen for neutral photodissociation. This yields energies $E_A$=4.6eV and $E_B$=3.8eV and notably $E_B$ is consistent with ($E_A$ - $E_{int}(I^*)$) from Equ.~{\ref{Equ_3}}, allowing that some energetic partitioning differences between the pathways is likely\footnote{for example, the data for CH$_3$I/C$_6$F$_6$  in Fig.~\ref{Fig_CH3I_C6F6_p_s} suggest that the CH$_3$ fragments in the I dissociation pathway have on average $\sim$0.24eV more internal energy than those following the I* pathway, which is a somewhat smaller difference than seen in the gas-phase data at this wavelength\cite{Eppink:1998ue}.}. The most parsimonious model would be that the photodissociation occurs via a charge-transfer complex formed between the extended half-bulk C$_6$H$_6$ layer and a CH$_3$I admolecule. This would involve photoexcitation promoting an electron from the C$_6$H$_6$ HOMO ($\pi$) band\cite{Sonoda:2007uz} to the CH$_3$I LUMO ($\sigma^*$). Schematically this process would be:


\begin{equation} \label{Equ_4}
\begin{split}
[C_6H_6]-&CH_3I\xrightarrow{h\nu}[C_6H_6]^+ -(CH_3I)^- \rightarrow \\
&\rightarrow \left\{ \begin{array}
{l@{\quad}l}
\rightarrow [C_6H_6]-I + CH_3 \\ \rightarrow [C_6H_6]-I^* + CH_3
\end{array} \right.
\end{split}
\end{equation}

where the [C$_6$H$_6$] denotes that it is the extended solid of C$_6$H$_6$ rather than a single molecule participant. The dissociation process with pathways corresponding to $I$ and $I^*$ outcomes as per Equ.~\ref{Equ_1} are consistent with the energy difference $E_A$-$E_B$ inferred. The observed intensities for the A and B peaks (Fig. {\ref{Fig_CH3I_thick_C6H6_p_s}}) would suggest that the two pathways in Equ. {\ref{Equ_4}} have comparable probabilities, similar to that of neutral photodissociation (e.g. Fig. {\ref{Fig_CH3I_C6F6_p_s}})\cite{DeBoer:1996df}. This scenario would have a single initial excitation energy, $E_A$, which is close to the electronic origin of the ${^1B_{2u}}$ band for benzene in the condensed phase at 90K, observed at 4.69eV\cite{Dawes:2017dpa}, possibly through the associated ${^1B_{2u}}$ exciton\cite{Bernstein:1968gs,Dawes:2017dpa}. Due to the ionic character of the charge-transfer complex, the energetics and lifetimes can be quite sensitive to the charge stabilization due to the local solvation environment\cite{Cheng:1995ur}. This might account for the differences seen in the TOF spectra for the thinner C$_6$H$_6$ films and explain why thicker films displayed the same features as seen for the 10ML C$_6$H$_6$ films. The only other system in the surface photochemistry literature that invoked an intermolecular charge-transfer mechanism for photodissociation, for CF$_3$I films on Ag(111), also noted that this mechanism was observed more prevalently for the thicker films\cite{Sun:1995vm}. 

An alternative explanation to the CT-complex would be that the dissociation outcomes for both the A and B features result in ground-state $I$ (upper pathway in Equ. {\ref{Equ_4}}), but that the dissociative energies come from quenching the lowest-lying triplet state excitations (here enhanced via the ``heavy-atom effect'' from CH$_3$I)\cite{DeBoer:1996td} that been noted in optical spectroscopy\cite{Dawes:2017dpa,Colson:1965aa} and electron-energy loss spectroscopy\cite{Swiderek:1998vq} studies of condensed benzene thin films. In this case, the quenching of the lower energy ${^3B_{1u}}$ state (origin energy estimated at $\sim$3.95eV\cite{Dawes:2017dpa} and $\sim$3.68eV\cite{Swiderek:1998vq}) would lead to peak B, while peak A would be due to quenching of the higher energy ${^3E_{1u}}$ state (origin energy $\sim$4.6eV\cite{Dawes:2017dpa}). It is harder to rationalize the similar intensities seen for the A and B features in Fig. {\ref{Fig_CH3I_thick_C6H6_p_s}} for this scenario.  

\subsection{ Photodissociation of Methyl Halides on \texorpdfstring{C$_6$F$_6$}{C6F6}}
\label{no_dea_in_c6f6}
A striking observation from the photodissociation data for the methyl halides on C$_6$F$_6$ thin films in the present work is the absence of dissociation via the CT-DEA pathway. Bond scission can be rapid via DEA due to the relative shift of the anionic potential energy surface in a condensed phase system, with large cross sections compared to the gas-phase of the same molecules. That photoelectrons of the requisite energies are generated in these C$_6$F$_6$ systems for the UV wavelengths used is expected based on workfunction measurements-- for C$_6$F$_6$ thin films on Cu(111) the workfunction is observed\footnote{for C$_6$F$_6$ on Cu(111), Ref.~{\onlinecite{Ishioka:2000ws}} reports $\Delta\Phi$=-0.39eV for the first layer, and a further $\Delta\Phi$=-0.03 for the second layer, and constant beyond; Ref.~{\onlinecite{Vondrak:1999ks}} reports $\Delta\Phi$=-0.16eV for the first layer and $\Delta\Phi$=-0.02eV for the second layer for the same system.} to decrease by $\sim$0.4eV. The bare surface workfunctions for the Cu(110) and Cu(100) substrates used in the present study are lower than for Cu(111) (Cu(110): 4.56eV; Cu(100): 4.73eV; Cu(111): 4.90eV)\cite{Derry:2015cza}, so although we do not know the workfunction for our systems of interest, we believe that it will be less than the photon energies used here (5.0eV and 6.4eV) so expect that both hot photoelectrons and low energy photoelectrons are present, such as that inferred by the CT-DEA observed for the methyl halides on C$_6$H$_6$ thin films.

The valence electronic structure of C$_6$F$_6$ thin films on metals has been studied by inverse photoemission\cite{Dudde:1990tc,Grass:1993dg} and two-photon photoemission\cite{Vondrak:1999ks,Gahl:2000eh,Ishioka:2000ws}. What appears to be of relevance for the current work is the C$_6$F$_6$ LUMO, which in the gas-phase is known for trapping low-energy electrons, while for thin films on metal surfaces, this $\sigma^*$ resonance has been found to form an extended quantum-well free-electron like state located between the Fermi and vacuum levels (roughly $E_F$+2.90eV, varying slightly with film thickness)\cite{Zhao:2014ia}. Two-photon photoemission has found that electrons populating this state have short lifetimes, varying from 7fs for 1ML and increasing to 32fs for 5ML films on Cu(111)\cite{Gahl:2000eh}. That this $\sigma^*$ state can quench surface electronic excitations has been found for the image states of C$_6$F$_6$/Cu(111) (energetically located just below $E_{vac}$, spatially located at the vacuum interface and having $\sigma$-symmetry), with anomalously lowered lifetimes (5fs for 2ML C$_6$F$_6$ and shorter for thicker films), apparently through decay and coupling to this LUMO-derived state\cite{Ishioka:2000ws}. The implication is that the energetically similar anionic states responsible for DEA in adsorbed methyl halides can couple effectively to this C$_6$F$_6$ extended state. This requires both the energetic alignment and also spatial arrangement so that the relevant wavefunctions allow rapid electron transfer to occur\cite{Fohlisch:2012bc}. It is understood that the state implicated in halomethane DEA in condensed systems is located at low energy, with substantial weight below the vacuum level, as photoinduced processes are often observed for photon energies less than the system workfunction. The methyl halide LUMO involved in DEA is a $\sigma^*$ molecular orbital, with significant weight on the halogen end of the molecule\cite{orbimol:2020}. As such, an oriented methyl halide will have spatial as well as energetic overlap for this state with the C$_6$F$_6$ layer below, creating favourable conditions for intermolecular quenching\cite{Zhou:1995}. Such a mechanism operating on the C$_6$F$_6$ thin film is analogous to the quenching seen for many photochemical systems when the molecules are adsorbed on or in close proximity to a metal or semiconductor surface, with an abundance of unoccupied valence states that can couple to the molecular excitation before dissociation proceeds. The property of this molecular thin film is novel in that CT-DEA is quite a rapid dissociation process, even compared to A-band neutral photodissociation  for the methyl halides, due to the short distance to the dissociative curve crossing. That the neutral photodissociation processes are not being quenched for the 2ML and thicker C$_6$F$_6$ films is evidently due to the lack of an efficient quenching pathway for these molecular excited states, regardless that the methyl halide LUMO is involved in both DEA and neutral photodissociation. That quenching of neutral photodissociation is observed for the 1ML film indicates that either the LUMO-derived C$_6$F$_6$ state shifted upward in energy is sufficient to couple with the neutral excitation or that the methyl halide excited state can couple with the Cu substrate states to facilitate quenching through the C$_6$F$_6$ monolayer. It is noteworthy that the observations here for methyl halides on C$_6$F$_6$ thin films are the obverse of findings for photodissociation of CH$_3$Br and CH$_3$Cl on CH$_3$OH thin films\cite{Jensen:2015dg}, where CT-DEA was enhanced but neutral photodissociation at $\lambda$=193nm was quenched, likely due to resonant hole-filling by the CH$_3$OH layer. This highlights the importance of the molecular film electronic structure and electron transfer in how the outcomes for excitations of co-adsorbed molecules can be altered\cite{Engelhart:2015dd}.

The degree to which the C$_6$F$_6$ LUMO band couples with the electronic states of the metal substrate has been a point of interest\cite{Zhao:2014ia,Fohlisch:2012bc} and for this reason we used both Cu(110) and Cu(100) substrates for this investigation, as the former has no surface bandgap between $E_F$ and $E_{vac}$ at the Brillouin Zone centre, while the latter does\cite{Grass:1993dg}. The comparisons of methyl halide photodissociation on C$_6$F$_6$/Cu(110) and C$_6$F$_6$/Cu(100) in the present study found no notable differences. This suggests that the methyl halide CT-DEA quenching is dominated by its interaction with the C$_6$F$_6$ adlayer and that any subsequent differences in how the C$_6$F$_6$ LUMO state couples to the underlying metal had no observable effect for the dynamics involved in CT-DEA of the methyl halides studied.

\section{Summary and Conclusions}
In examining the photodissociation of methyl halides adsorbed on thin films of C$_6$H$_6$ and C$_6$F$_6$, two observations were made that highlight the significance of details in electronic structure for these molecular solids for photodissociation processes. The photodissociation mechanisms found for CH$_3$Cl and CH$_3$Br on C$_6$H$_6$ thin films on Cu substrates were consistent with neutral and CT-DEA processes, while the photodissociation of CH$_3$I adsorbed on this film displayed a new dissociation pathway not seen for the other methyl halides and is apparently due to the specifics of the heteromolecular  CH$_3$I/C$_6$H$_6$ system, likely via a charge-transfer complex excitation, conceptually an intermediate case between the neutral photodissociation and DEA mechanisms.

The photochemistry and photodissociation of halomethane molecules on metal and semiconducting surfaces have been studied extensively in the past 30 years and one of the common features in much of that work has been the ``redshifted'' photochemistry observed in various contexts, which has generally been ascribed to hot photoelectrons and/or photoelectrons generated at the substrate and transmitted to the adsorbed halomethane. For the methyl halides adsorbed on thin films of C$_6$F$_6$/Cu, neutral photodissociation is seen that is consistent with expectations, while the CT-DEA process is entirely quenched. The selective quenching of CT-DEA is a striking finding as the usual expectation from the surface photochemistry literature is that CT-DEA can persist when neutral photodissociation is quenched-- here we find the opposite being the case. This is due to an efficient quenching mechanism through a low-energy unoccupied valence band formed in the C$_6$F$_6$ thin film.


%
%

%


\bibliography{C6H6_C6F6_Paper_Bibliography}

\end{document}